\documentclass[epj]{svjour}
\usepackage{graphicx}
\usepackage{epsfig}
\usepackage{amsmath}
\usepackage{amssymb}
\usepackage{amsfonts}
\usepackage[utf8]{inputenc}
\usepackage{hyperref}
\usepackage[hyphenbreaks]{breakurl}
\usepackage{color}

\begin{document}
\title{Compelled to do the right thing}

\author{
M. F. Laguna\inst{1,}\thanks{lagunaf@cab.cnea.gov.ar}
\and
G. Abramson\inst{1,}\inst{2,}\thanks{abramson@cab.cnea.gov.ar}
\and
J. R. Iglesias\inst{3,}\inst{4,}\thanks{roberto@if.ufrgs.br}
}

\institute{CONICET and Centro At\'omico Bariloche, R8402AGP Bariloche, Río Negro Argentina
\and
Instituto Balseiro, R8402AGP Bariloche, Río Negro, Argentina
\and
Instituto de F\'{\i}sica, UFRGS and Instituto Nacional de
Ci\^encia e Tecnologia de Sistemas Complexos (INCTSC), Caixa Postal 15051,
91501-970 - Porto Alegre - RS, Brasil
\and Programa de P\'os-Gradua\c{c}\~ao em Economia, UFRGS, Av. Jo\~ao Pessoa 52, 90040-000 - Porto Alegre - RS, Brasil}

\abstract{We use a model of opinion formation to study the consequences of some mechanisms attempting to enforce the right behaviour in a society. We start from a model where the possible choices are not equivalent (such is the case when the agents decide to comply or not with a law) and where an imitation mechanism allow the agents to change their behaviour based on the influence of a group of partners. In addition, we consider the existence of two social constraints: a) an external authority, called monitor, that imposes the correct behaviour with infinite persuasion and b) an educated group of agents that act upon their fellows but never change their own opinion, i.e., they exhibit infinite adamancy. We determine the minimum number of monitors to induce an effective change in the behaviour of the social group, and the size of the educated group that produces the same effect. Also, we compare the results for the cases of random social interactions and agents placed on a network. We have verified that a small number of monitors are enough to change the behaviour of the society. This also happens with a relatively small educated group in the case of random interactions.}



\maketitle

\section{Introduction and review of the original model}
\label{intro}
The problem of opinion formation has been studied from several perspectives, including quantitative approaches, by many authors in the last decades. The seminal ideas of Axelrod on the dissemination of culture \cite{axelrod} have been re-ellaborated from many interdisciplinary sides. Of particular interest are also the fundamental papers of Galam (see for example \cite{galam82,galam86}, and also the recent review \cite{galamrev}). In many of these works, the authors propose simple quantitative models that try to capture the fundamental processes determining the emergency of consensus in a population (see \cite{castellano} and \cite{vicky} for thorough reviews). Example of this situation are voter models, in which the dynamics is analyzed using the tools of the statistical mechanics of magnetic systems \cite{mobilia03,mobilia07,lambiotte08}. 

It is generally assumed that the choice of the agents is between equivalent options. There are notable exceptions, such as \cite{galam86} cited above. Indeed, in many social situations the options presented to the individual may not be equivalent. For example, each agent may have a personal bias towards one or the other, as studied in \cite{galam91,galam97}. In other cases, the options may be endowed with an intrinsic value derived from cultural, ethical or legal foundations. In a recent article we studied a simplified instance of such a problem~\cite{paper01}. Possible examples of widespread---however il\-le\-gal---conducts are the decision of using a cell phone while driving a car, of not stopping at a red light, or of driving while intoxicated. Beside the subject's idiosyncrasy, the choice of one attitude or the opposite is influenced (and even determined) by their background (education, cultural values, etc.) as well as the pressure of the social environment \cite{mirta} and of the authorities \cite{authority}.

From a psychological point of view, the influence of a social group depends on the size of the group, on their persuasion capability, and on the distance from the subject (which may be spatial or abstract, in the sense of personal relationships) \cite{lantane1981}.
The resistance of a society to the convergence of its members' opinions has been considered by ourselves in~\cite{laguna2004}, and also in \cite{galam07,porfiri2007,mobilia07}. Besides the influence of social acquaintances, in real societies other factors may be of relevance: authorities can introduce a system of control and punishment~\cite{crimeandpunishment} through a municipal or road police or even, in some social events, by guard duties or other security officials. This has been studied in \cite{kacperski}, where a strong leader and external social impact are present. In addition to this, there may exist a group of more educated and responsible people that could influence the rest of the society but will not be influenced by them. This group forms effectively a core of the ``correct'' opinion or behaviour.

In this paper we study the consequences of these mechanisms that try to ``constrain'' the society to adopt the correct behaviour (which will also be referred to as ``opinion''). As in Ref.~\cite{paper01}, we consider a system where the agents may adopt one of two options with different values. The model of social influence is implemented as an imitation behaviour: agents tend to imitate the opinion of the others, which may be the right or the wrong one. As a result of this social influence agents may change their opinion. In addition to this, we consider also the following two new ingredients. On the one hand the effect of authority, which we already addresed in \cite{authority}. Here we define agents or mechanisms called ``monitors'', who are in some way external to the social group and have the effect of imposing the correct opinion when the social agent is in contact with them. In real systems they may act by exerting some kind of controlling or punitive action: cameras, admonitions, reprimands or fines. Many times the presence of the monitor acts as a dissuasive factor---and the eventual fine too---enforcing the right behaviour on the agents that come to interact with them. An example of monitors are traffic police staying at a crossroads and applying fines to drivers violating the red light or using their cell phones; other examples are school monitors or surveillance cameras in public places.

The second ingredient is the effect of an educated group that never abandons the correct opinion. The effect of this core group is the same as the conventional social influence by imitation. Indeed, the existence of inflexible agents has been considered before in \cite{galam07,galam91,galam97}. The fact that they cannot be induced to adopt the ``wrong'' opinion produces a different dynamics, as was shown in \cite{galam10} in the context of the influenza pandemics. Indeed, we will show that, in many cases, it is possible to induce a major change in the opinion of the population by setting a fraction of ``polite'' or ``educated'' agents (those supporting the ``right'' option) as well as through controls. Both influences are able to generate a change in the society in a reasonable period of time. The fraction of educated agents necessary to induce a global change depends on the relative value of the opinions, on the average adamancy of the group and on the connectivity of the network.

Let us briefly review the dynamics of the original model presented in Ref.~\cite{paper01}. We consider a population of $N$ interacting agents, each of them having an {\it opinion} attribute, $x$, and a tendency to conserve this opinion, that we call {\it adamancy}, $a$. The opinion is bi-valuated, $x \in \{x_1,x_2\}$, with $x_2>x_1$, implying that opinion 2 is more valuable than opinion 1.
The dynamics of the system is as follows. At each time step an agent $i$ is
picked up at random. This agent is prompted to change his opinion because
of the influence of a group of $k$ individuals. These are chosen at random at each interaction and a poll is performed:
\begin{equation}
\text{If: } a_i x_i +n_i x_i-(k-n_i)x'_i \begin{cases} \geq 0 & \text{then agent $i$ keeps} \\
&\text{opinion $x_i$},\\
<0 & \text{then agent $i$ changes}\\
& \text{to opinion $x'_i$,}
\end{cases}\label{compar}
\end{equation}
where $n_i$ is the number of agents (belonging to the influence
group of size $k$) sharing the same opinion as agent $i$ and $x'_i$ is the
alternative opinion, sustained by the remaining $k-n_i$ agents.
Thus, there is a competition between the two opinion values. In addition, note that agent $i$
weighs his own opinion (and only this one) with his adamancy $a_i$.
The final state depends on the distribution of adamancy $a$ of the population, on the fraction $f_h$ of agents that have initially the higher opinion, and also on the distribution of $k$. The relative values of $x_1$ and $x_2$ are also relevant, but will be kept fixed in our study at $x_1=1$ and $x_2=2$ (as in \cite{paper01}).This dynamics has only two attractors in a strict sense: consensus of $x_1$ or consensus of $x_2$. Nevertheless, the time needed to reach such stable states can be extremely long and irrelevant in any social context. Typical phase diagrams will be shown in the left panels of Figs.~\ref{poli-delta} and ~\ref{poli-exp}. From Eq. (\ref{compar}) it is clear that for $a/k>2$ the agents cannot modify their opinions as a result of the interactions. The consequence in the diagram (Fig. \ref{poli-delta}, left panel) is a coexistence of opinions for all values of $f_h$, reflecting the fact that no interaction ends up in a change of opinion. This region represents a very stubborn society, and the system is frozen at the initial condition. See Ref. \cite{paper01} for a detailed explanation of the original model.

In the following sections we build upon this model and discuss the extended results.

\section{Enforcing the right behaviour}
\label{res}

Let us now consider two different mechanisms that can modify the time evolution and the attractors of the dynamics. These mechanisms are aimed to improve the chances of adoption of the higher valued opinion. The first one is to introduce monitors, who induce the agents to adopt the desired opinion. We model these monitoring agents as external to the society and having infinite persuasion. This means that when agents with opinion $x_1$ detect, or are detected by, a monitor, they always switch to opinion $x_2$, independently of their adamancy and the opinion poll of the group of influence. This change is not permanent, and the agent is able to return to $x_1$ if the social pressure, given by Eq.~(\ref{compar}), makes it so in subsequent interactions.

A second modification in the dynamics is the existence of a ``core'' of educated agents, i.e. agents that are---or have been---convinced of the benefits of the higher opinion and that have an infinite adamancy. When interacting with other agents they never change their opinion but, unlike the monitors, their influence on other agents is the same as that of any other agent with opinion $x_2$. This core of educated agents provide an unsupervised pressure favouring the higher valued opinion $x_2$.

In this section we detail the effects of these two types of social influence within a mean field scheme, while in Section~\ref{redes} we discuss the situation where the agents are distributed on a complex network.

\subsection{Monitors}

To start with and following Ref.~\cite{paper01}, we consider a system of $N=1000$ agents with an initial fraction $f_h$ of them sustaining the right opinion ($0 \le f_h \le 1$).  We call $P$ the probability of an agent to be detected by a monitor. If the agent is detected, he changes his opinion to $x_2$, otherwise he interacts with $k$ neighbours and perform the comparison given by Eq.~(\ref{compar}). As a result of the poll he can change or not his state. The simulation runs for a total time of 100 interactions per agent. We consider this a reasonable period of time to study social changes (even though in some social situations more interactions might be relevant for the decision making; see ~\cite{paper01} for an analysis on this dependence). We remark that we will study two situations: a) a fixed number of neighbours (particularly we will consider a delta probability distribution with $k=10$), and b) a number of neighbours determined at random from an exponential probability distribution (with $\langle k\rangle =10$). Note that in this latter case it is more likely to have a low number of connections that a high one.

\begin{figure}[t]
\centering
\includegraphics[width=\columnwidth]{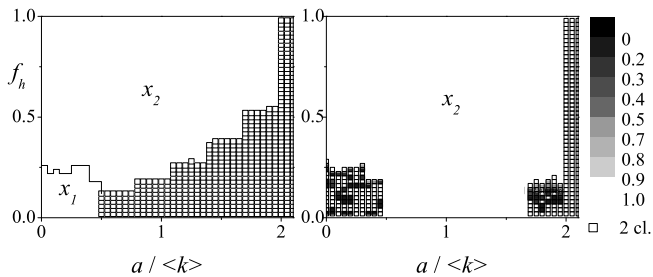}
\caption{Phase diagrams showing the effect of the monitors in a population of $N=1000$ agents, after $100$ interactions per agent (in average), and homogeneous connectivity with $k=10$. Left panel: no monitors~\cite{paper01}. Right panel: $1$ monitor per $100$ agents ($P=0.01$). The region where the society converges to a state of opinion $x_2$ is much wider in this last case than in the situation with no monitors. In both diagrams, each point consist of $100$ realizations. White squares correspond to states of coexistence of opinions (named 2 clusters states, in accordance with~\cite{paper01}). The scale of grays indicates the fraction of realizations that arrive to a consensus (with darker indicating rare consensus, lighter frequent consensus. White regions (without squares) correspond to consensus of either opinion, as shown.} \label{poli-delta}
\end{figure}

We have verified that the dynamics is very sensitive to the action of the monitors. In Fig.~\ref{poli-delta} we compare the phase diagram obtained with no monitors (left panel), with the diagram obtained with a probability $P=0.01$ of being monitored, i.e., there is in average $1$ monitor for every $100$ agents (right panel). It can be observed that the area occupied by the consensus of the opinion $x_2$ in the right panel is much bigger than in the unsupervised case (left panel). The region that contributes most to the increase of consensus of the higher opinion is the one with $a/\langle k\rangle <2$, a boundary that depends on the relative values of $x_2$ and $x_1$ in the present model. When $a/\langle k\rangle >2$ the system converges to opinion $x_2$ but over very long times, because of the diluted interaction with the monitor. Very long times are not realistic in social sciences, but we should remark that even if both opinions coexist we have verified that the number of agents sustaining opinion $x_2$ constitute a very large majority.

Besides, the original model had a small region of consensus of opinion $x_1$ in the region of low adamancy and low $f_h$ that has disappeared. This region exhibits now a mixed state, produced by two competing effects. On one side, monitors convert agents to opinion $x_2$. On the other, as their adamancy is low, agents are quickly convinced by their neighbours to choose $x_1$. This dynamics leads to a fluctuating state that is perpetuated over time, and no consensus is reached.

In short, if we consider an infinite time of interaction (which is not reasonable for social systems) the system would converge to the opinion $x_2$ for any value of the adamancy, with the exception of the regions described as fluctuating states in the previous paragraph.

\begin{figure}[t]
\centering
\includegraphics[width=\columnwidth,clip=true]{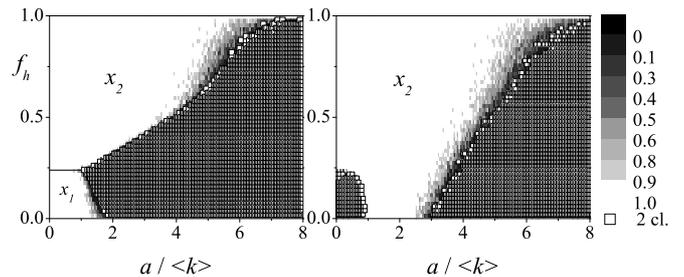}
\caption{Same phase diagram of Fig. \ref{poli-delta} but with an exponential connectivity distribution with $\langle k\rangle=10$. Again, the left panel corresponds to the situation in which no monitors are present~\cite{paper01}. The right panel shows the diagram obtained after the inclusion of $1$ monitor per $100$ agents. Note that the horizontal scale is not the same of Fig. \ref{poli-delta}.} \label{poli-exp}
\end{figure}

For the sake of completeness we have also studied the case of an exponential distribution of connectivity, with $\langle k\rangle =10$ (see Fig.~\ref{poli-exp}). It can be verified that the region of consensus of opinion $x_1$ disappears, but the area covered by the higher opinion, $x_2$, even if greater than in the case with no monitors, does not exhibit a significant variation. This occurs because the exponential distribution privileges lower connectivities, becoming harder to reach consensus of the high valued opinion, even with the same concentration of monitors.

\begin{figure}[b]
\centering
\includegraphics[width=0.8\columnwidth,clip=true]{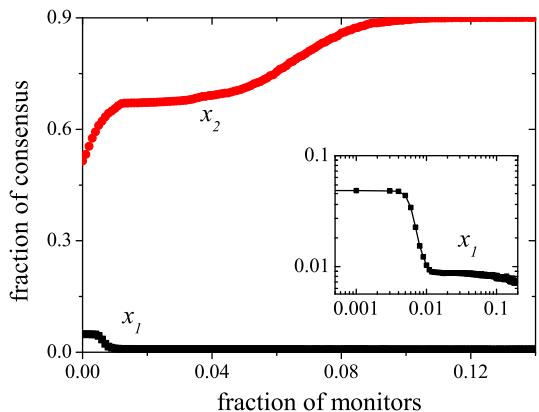}
\caption{Effect of the relative number of monitors for a delta distribution of connectivities. Main panel: Fractions of the total area for which the consensus of opinions $x_1$ and $x_2$ are obtained (after $100$ interactions per agent in average) as a function of the fraction of monitors, $P$. It is possible to see that the area covered by the opinion $x_1$ goes to zero for a very small number of monitors, 1 per 100 agents ($0.01$ on the abscissa). Besides, the opinion $x_2$ stabilizes at $90\%$ of the total area for 1 monitor per 10 agents ($0.1$ on the abscissa). Inset: Area of $x_1$-consensus in logarithmic scale.} \label{poli-x2x1}
\end{figure}

If this is the effect of just ten monitors (representing 1\% of the population), one can assume that increasing their number will induce full adoption of opinion $x_2$ for any value of the adamancy. One possible way of verifying this point is to calculate from the phase diagrams the relative areas of the regions where the opinions $x_1$ and $x_2$ are stable, compared with the area of the rectangle going up from $ 0 \le f_h \le 1$ and adamancy from $0 \le a \le 2.5$.
These results are presented in Fig.~\ref{poli-x2x1} for the case of the delta distribution of connectivities. The relative areas of the $x_1$ and $x_2$ opinions are plotted as a function of the probability $P$, and it is possible to see that a very small number of monitors is enough to induce an almost full consensus on the population. This happens in two stages. When there is just one monitor every $100$ agents, as is the case in the diagram of Fig. ~\ref{poli-delta}, there are almost no more realizations with consensus of $x_1$ (it is strictly zero when the number of monitors is greater than $2 \%$ of the population). With ten times as many monitors the population reaches consensus of $x_2$ in $90\%$ of the area considered. A smooth transition consisting in an increasing number of $x_2$ opinions connects these two values. In this region, the growing presence of monitors speeds up the dynamics. Accordingly, the $x_2$ area grows even if the number of interactions per agent is the same in all cases. When $P=0.1$, the $x_2$ curve reaches $90\%$ of the total area and stops growing. For this density of monitors, all coexistence regions of $a/\langle k\rangle <2$ (except the fluctuating ones) are converted to $x_2$-consensus.  In order to convert populations with $a/\langle k\rangle >2$ and reach maximum consensus of opinion $x_2$, it would be necessary more than $1,000$ interactions per agent, a number ten times greater that the one considered to be reasonable in the present work.

\subsection{Social pressure}

The effect of a social ``hard core'' group in the society is also impressive. Let us consider that a fraction of agents $n/N$ stand by the right opinion $x_2$ and have an infinite adamancy. The obtained effect is similar to the one of including monitors, but with a sensible difference: a few agents with infinite persuasion induce a change in the society much faster than a group of agents with infinite adamancy. Adamancy guarantees those agents will not change opinion, but their ability to persuade is not changed by this factor. As a consequence, a bigger group of social pressure is necessary to change the behaviour of the society.

\begin{figure}[t]
\centering
\includegraphics[width=\columnwidth,clip=true]{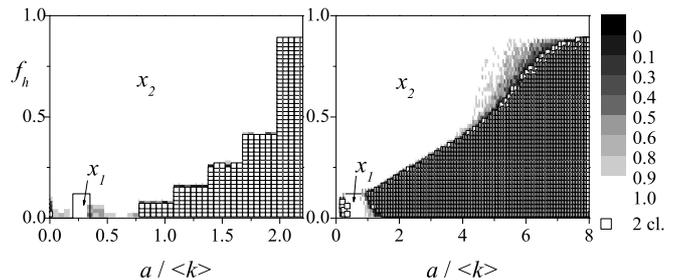}
\caption{Effect of $100$ educated agents in a population of $N=1000$ agents, with $\langle k\rangle=10$ and a final time of $100$ steps per agent (in average). Observe that the educated agents cannot change their opinions. So, $f_h=0$ implies that no regular agents start with opinion $x=2$, but a 10\% of educated agents are already present in the whole system. Left panel: delta distribution of connectivities. Right panel: exponential distribution of connectivities. Comparing with the left panels of Figs.~\ref{poli-delta} and ~\ref{poli-exp}, one can verify that the region where the society converges to a state of opinion $x_2$ is wider and goes to lower values of $f_h$, than in the situation with no educated agents.}
\label{nucleo100}
\end{figure}

Let us first look at the phase diagram, comparing the case with no monitors and no ``hard core'' represented on the left panels of Figs. \ref{poli-delta} and \ref{poli-exp}, with the case of $10 \%$ of the population making part of the ``hard core'' group. These results are shown in Fig.~\ref{nucleo100}. There is certainly an increase of the region where opinion $x_2$ is the consensus but the size is not very different when compared with the original results. This means that the region of consensus of the high valued opinion $x_2$ is larger when the ``hard core'' is in action, but not as large as when there are monitors. The paradoxical result is that the region of stability of the lower opinion is shifted from adamancy equal zero to regions with finite adamancy around $a/\langle k\rangle =0.25$ for the delta connectivity distribution and $a/\langle k\rangle =0.5$ for the exponential one. On the other hand, for very low values of the adamancy a strip in the area previously occupied by the lower opinion, becomes now a region of not defined consensus.

\begin{figure}[b]
\centering
\includegraphics[width=0.8\columnwidth,clip=true]{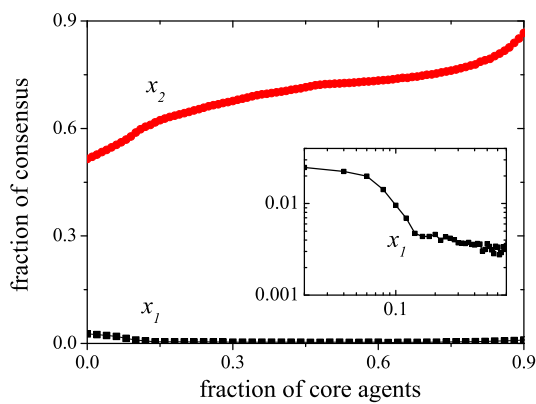}
\caption{Effect of the relative number of educated agents for a delta distribution of connectivities. Main panel: Fractions of the total area for which the consensus of opinions $x_1$ and $x_2$ are obtained (after $100$ interactions per agent in average) as a function of the fraction of core agents. Inset: Area of $x_1$-consensus in logarithmic scale. Note that the area of $x_1$-consensus goes to zero when the size of the hard core group is of the order of $12 \%$ of the total population.} \label{nucleo-21x1}
\end{figure}

This effect can be verified by doing the same kind of graphic as the one of Fig.~\ref{poli-x2x1}. We have represented in Fig.~\ref{nucleo-21x1}  the area of consensus of both opinions as a function of the fraction of agents that participate in the ``hard core'' group of influence. More than $10\%$ of the population taking part in the ``hard core'' nucleus is necessary in order to make the area of the $x_1$ opinion go to zero. Moreover, the increase of the area with $x_2$ consensus increases much more slowly than in the case with monitors and arrives to cover $75\%$ of the total area when the hard core nucleus is of the order of $70\%$ of the agents.

From these results it can be concluded that infinite persuasion is much more effective than infinite adamancy. We can speculate that it is easier to put $1 \%$ monitors in the society than to educate a group of the order of $15\%$ of the population. Certainly, in such a case there could be a social cost to be considered. Police or highway patrols need to be trained and must be paid. Besides, an effective $1\%$ ``presence'' could imply a higher number of agents because of timetables, localization, etc. Also, in some cases they could behave in ways different from the ``perfect'' monitor we have imagined here: incorruptible agents with infinite persuasion. On the other hand to educate a hard nucleus may be easier and cheaper, with a sustained effect and also a less traumatic one: it is always preferable to convince people with arguments than with fines.

\begin{figure}[t]
\centering
\includegraphics[width=0.8\columnwidth,clip=true]{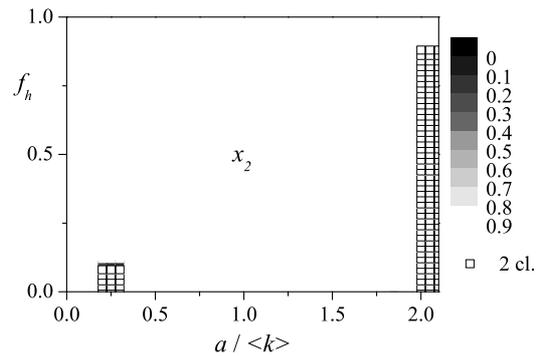}
\caption{Phase diagram with $P=0.01$ and $100$ hard core agents, in a system with $1000$ agents. Note that the effects of monitors and hard core nucleus are mutually magnified.}
\label{polinuc-CM}
\end{figure}

Nevertheless, in real societies both kinds of compelling factors are present, and to analyse a more realistic situation we have performed a simulation where both, monitors and a hard nucleus, are present. In Fig.~\ref{polinuc-CM} we show the diagram obtained for the case of $100$ hard core agents and $P=0.01$, which represents one monitor every $100$ agents. Although it could be assumed that monitors would mask the presence of the nuclei, we see that this is not the case, since both effects are enhanced. Moreover, the region of $x_1$-consensus vanished and the higher consensus occupies almost the entire available area.

\section{Complex social networks}
\label{redes}

The previous results, as well as the ones of Ref.~\cite{paper01}, were obtained within a mean field scheme. This means that at each time step an agent $i$ is randomly selected and can interact with any other in the system.
But a question arises: How do the phase diagrams change if the underlying structure of the population is different? In other words, what happens if the social influence comes from a social nucleus of close acquaintances and not from a random chosen group? The effect of the structure and dynamics of complex networks in the process of opinion formation has been extensively investigated in recent years. For example, Grabowski and Kosinski have analyzed an Ising model in which agents are endowed of a authority parameter and belong to a complex network \cite{grabowski}. In this section we analyse the case in which the agents are the nodes of a complex network with a given topology. Thus, agent $i$ interacts only with his first neighbours in the network.

We explore three archetypical complex networks: a) the Strogatz-Watts small-world (SW) network, b) the Erd\"os-R\'enyi (ER) random network (also known as Poisson network), and c) the Barabási–Albert (BA) scale-free network \cite{watts,strogatz,newman,barabasi}.
Small-world networks are particularly interesting, since they lie at some intermediate point between fully ordered and fully random networks. They possess a large degree of local clustering and a relatively small minimal path length connecting nodes throughout the system. To construct a SW network we apply the usual procedure~\cite{watts}: we start with a regular ring where each node has connectivity $k$ and  reconnect links at random with probability $\beta$.  Varying $\beta$ makes it possible to evolve from a regular lattice ($\beta=0$) to a random graph ($\beta=1$). Intermediate values of $\beta$ generate the so-called {\it small world} networks. The Erd\"os-R\'enyi topology corresponds to a Poissonian distribution of connections with mean $\langle k\rangle=K$: $P(k) = e^{-K} K^k/ k!$. Finally, scale-free networks~\cite{barabasi} represent systems where there is a wide distribution of connectivities and are characterized by the presence of a few ``hubs'' or elements
with an extremely high connectivity. These hubs hold the network together and produce a very small minimal path length. Also, the distribution of the degree follows a power law $P(k) \sim k^{-n}$~\cite{barabasi}.

For the three networks studied we consider an average connectivity $\langle k\rangle=10$. Moreover, for the SW network we fix the value of the parameter $\beta = 0.1$ and for the BA network the exponent is $n=3$. All the results of this section correspond to a system of $1000$ agents and the results are obtained after $100$ interactions per agent (in average). Each point of the phase diagrams shown below is the result of $10$ different networks with identical parameters.

\subsection{Complex networks and monitors}

\begin{figure}[t]
\centering
\includegraphics[width=\columnwidth,clip=true]{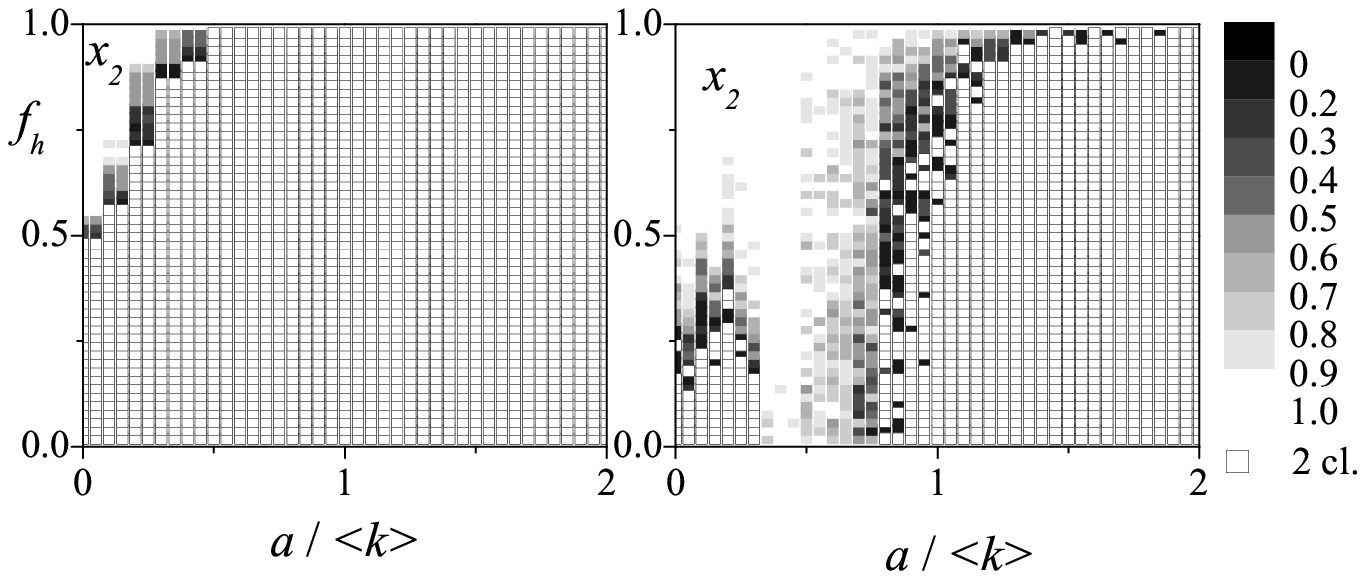}
\includegraphics[width=\columnwidth,clip=true]{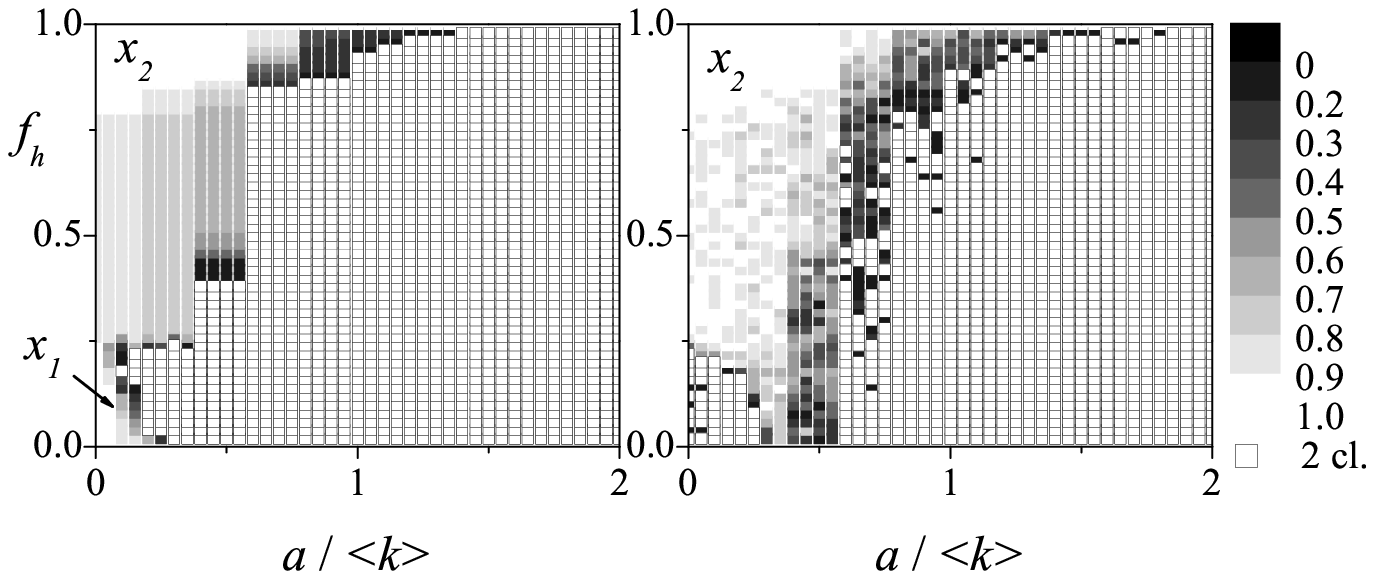}
\includegraphics[width=\columnwidth,clip=true]{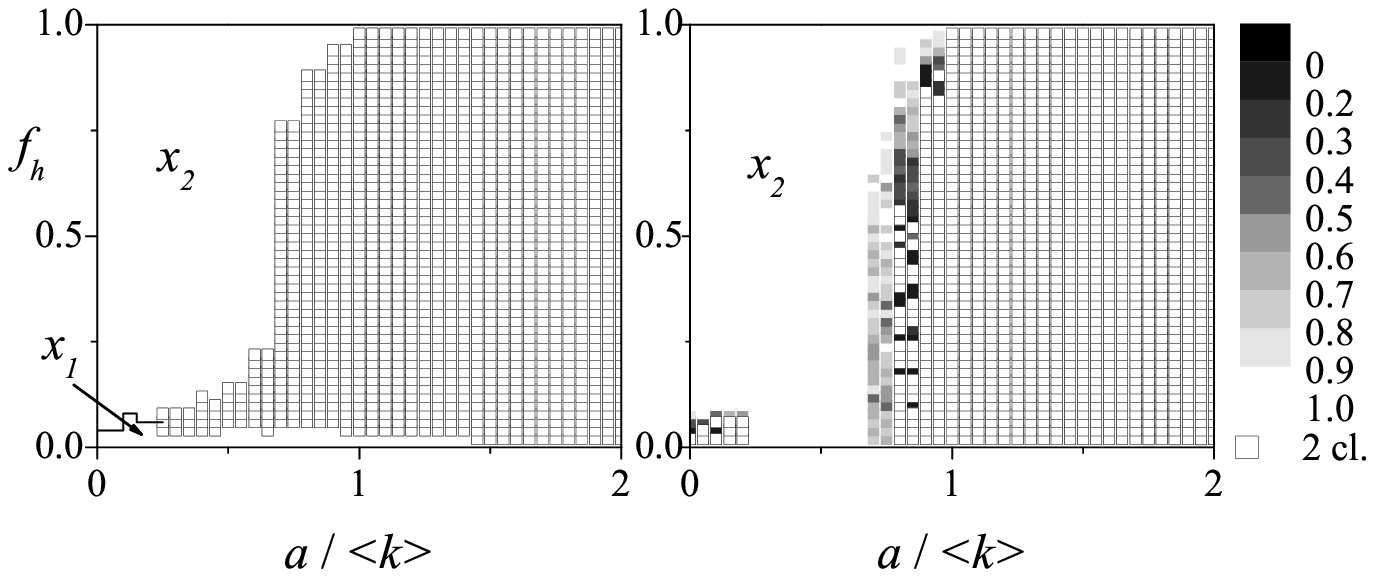}
\caption{Phase diagrams of networks, with free agents and with monitors. Top: SW network. Center: ER network. Bottom: BA network. Left panels: neither monitors or hard core nucleus. Right panels: one monitor every $100$ agents ($P=0.01$). The meaning of the gray scale and symbols is the same as in the previous figures.} \label{redes-poli}
\end{figure}

For sake of clarity, we start this section by analysing to what extent the original phase diagram (Fig.~\ref{poli-delta}, left panel) changes if we put the agents on the nodes of a complex network. This is shown in the left panels of Fig.~\ref{redes-poli} for the SW, ER and BA networks (top, center and bottom, respectively). In all cases the most remarkable difference is the reduction of the area of $x_2$-consensus, indicating that networks are less effective to propagate the right behaviour than random interactions. This fact can be understood if we remember the mechanism of selection in the mean field approach: each time that an agent is chosen the $k$ partners are different (because they are picked at random). This seems to be more effective to propagate information (or, in this case, an opinion) than any static network. We also verify that consensus is reached in networks, but at times much larger than in the mean field case.
However, a significative difference between the networks is that for the SW one the region of stability of the lower valued opinion disappears. This means that networks are inefficient to propagate both opinions, and worse for the lower valued one. Random (ER) or BA networks exhibit larger regions of stability for both opinions, preserving a very small region of stability of opinion $x_1$.

Now, we include in the dynamics the effect of monitors, keeping in mind that this means that each agent has a probability $P$ of interacting with one monitor, but the monitors are ``flying agents,'' not fixed in the network. In this situation and for one monitor every $100$ agents one observes in Fig.~\ref{redes-poli} that the area of $x_1$-consensus disappears, whereas the $x_2$-consensus area grows, although not as much as in the mean field case.

\subsection{Social pressure in complex networks}

\begin{figure}[t]
\centering
\includegraphics[width=\columnwidth,clip=true]{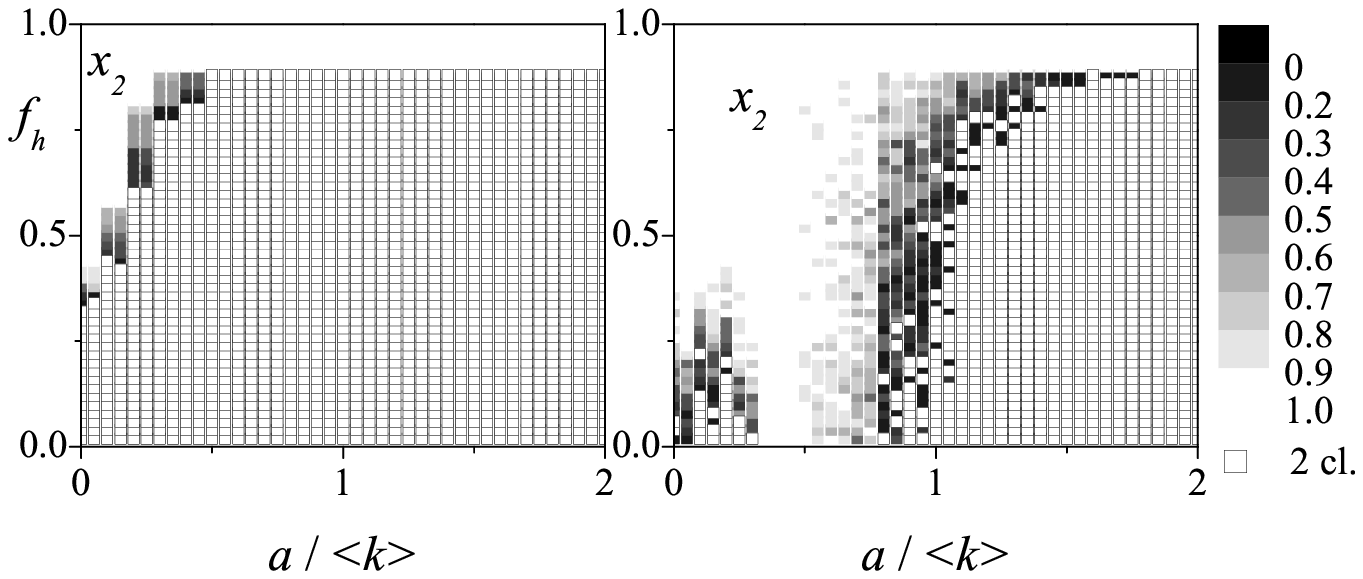}
\includegraphics[width=\columnwidth,clip=true]{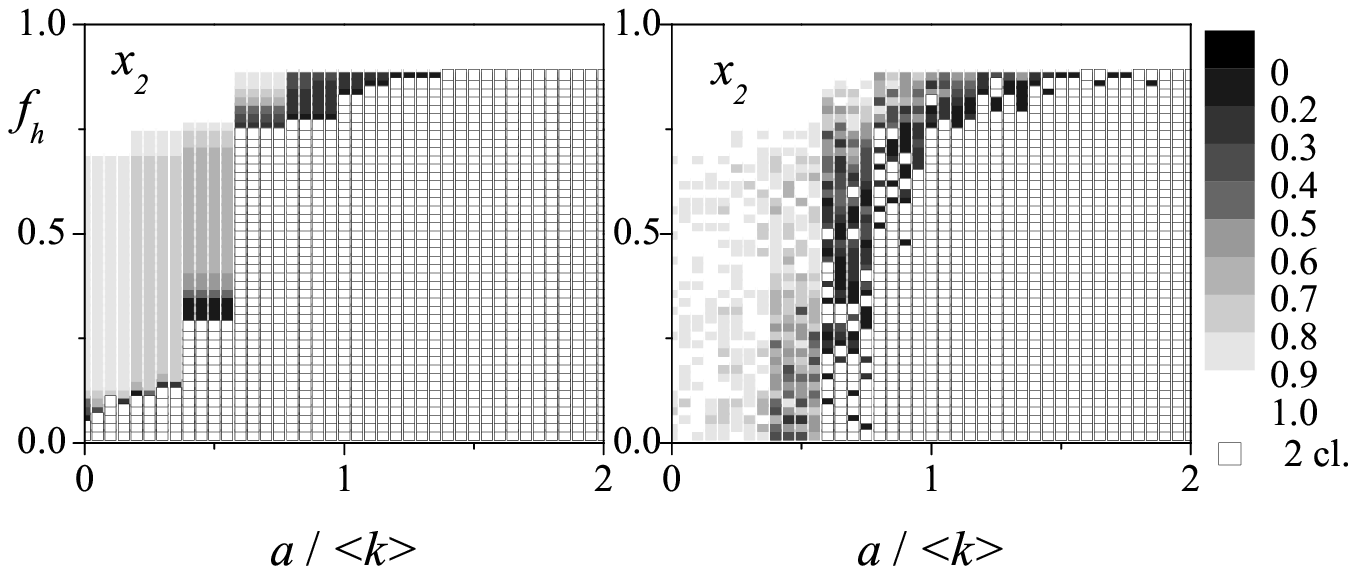}
\includegraphics[width=\columnwidth,clip=true]{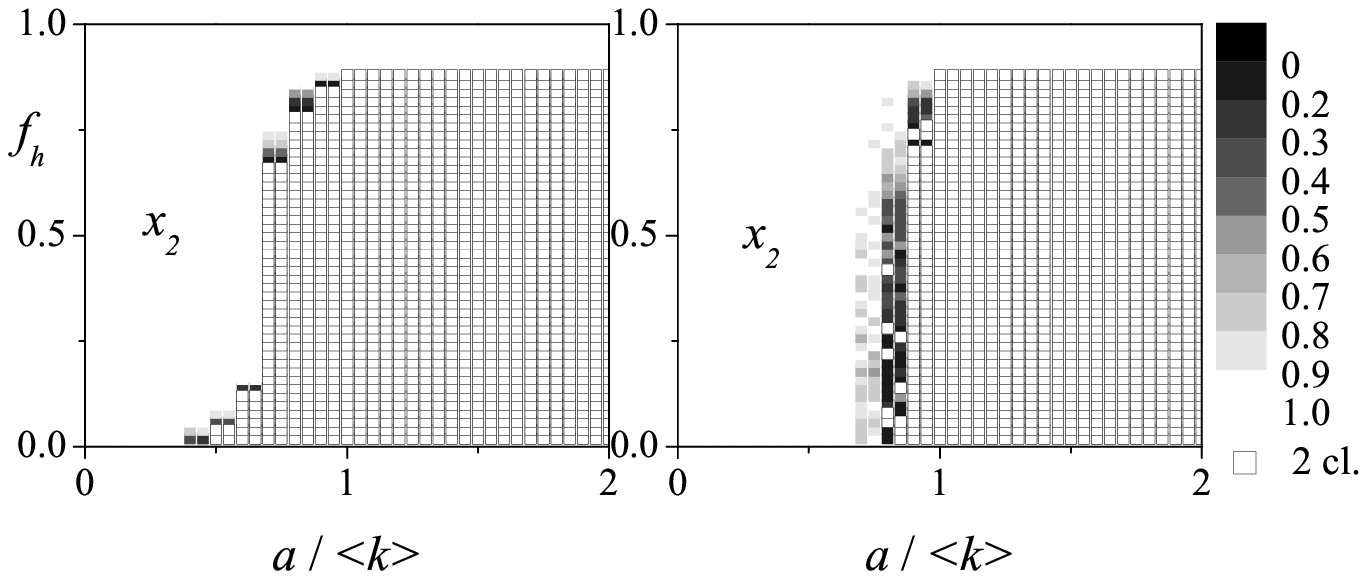}
\caption{Phase diagrams of networks, with an educated nucleus and monitors. Top: SW network. Center: ER network. Bottom: BA network. Left panels: just $100$ hard core agents. Right panels: $100$ hard core agents and $1$ monitor every $100$ agents ($P=0.01$).}
\label{redes-nuc}
\end{figure}

In this section we study the effect of a hard core group inside the complex network. In the left panels of Fig.~\ref{redes-nuc} we show the case in which a nucleus of $100$ agents is included in a society composed by $1000$ agents. Although the region of $x_2$-consensus grows compared with the case without monitors or nucleus (left panels of Fig.~\ref{redes-poli}), the effect of the educated nucleus seems to be much less significant than in the case with monitors.

For the sake of completeness we also include the results for the most realistic case of having together monitors ($P=0.01$) and a group of $100$ educated agents. The corresponding cases for the three model networks are shown in the right panels of Fig.~\ref{redes-nuc}.

\begin{figure}[t]
\centering
\includegraphics[width=\columnwidth,clip=true]{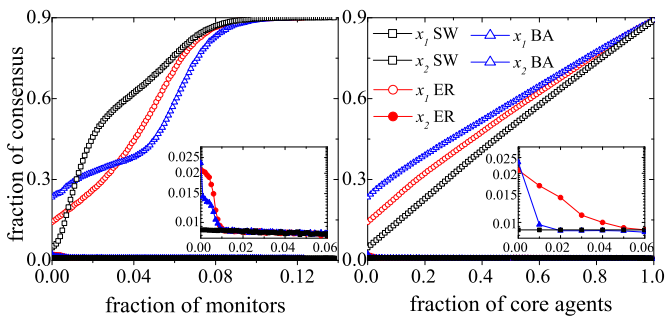}
\caption{Effect of the relative number of monitors (left panel) and educated agents (right panel) when they are distributed on three different social networks. The plots show the relative areas of the two consensus regions. Note that the scale is the same for the vertical axis in both graphs, but is different for the two abscissas.}
\label{x2x1-polired}
\end{figure}

Finally, we present in Fig.~\ref{x2x1-polired} the relative areas of the $x_1$ and $x_2$ opinions for the three complex networks. In the  left panel we plot this as a function of the concentration of monitors, $P$, whereas in the right panel we use the fraction of agents participating in the hard core. A very small number of monitors is, again, very effective to induce a change in the population. As in the mean field model, just 1 monitor every $100$ agents is enough to suppress the consensus of $x_1$, and ten times as many are necessary to reach consensus of $x_2$. This reflects the fact that monitors are outside of the network. The intermediate region, of neither consensus of $x_1$ nor $x_2$, instead, shows differences between the networks: the SW model is more effective in the dissemination of $x_2$, and the BA model is the worst. We conjecture that the short-cut links characteristic of the former are responsible for this advantage, while the hubs of the latter can hinder the process. On the other hand the educated hard core only produces consensus of $x_2$ for very high (and unrealistic) concentrations. Nevertheless, the consensus of $x_1$ effectively disappears for much smaller cores (see the inset of Fig.~\ref{x2x1-polired}, right panel). This effect is achieved with a minimal core in the case of the SW model. Both the BA and the ER graphs require bigger ones, albeit also rather small: about 1\% and 5\% of the population respectively.

It is also interesting to remark that the SW network is more efficient to optimize the work of the monitors, while it is the less efficient in the case of action of the hard core nucleus. On the other hand, the opposite result is observed for the BA network. Moreover, the structure of the networks is clear when one considers the hard core nucleus, but in the case of monitors, if $P > 0.1$ the details of the network become irrelevant.

In conclusion, if the agents are on a (not evolving) network, the effect of monitors is very similar to the mean field situation, and this is so because the monitors are ``outside'' of the network. On the other hand, when educated groups are introduced, the dynamics of opinions is very slow and the change to the appropriate behaviour is almost impossible for a reasonable number of members of the educated group. The fact that the educated group is distributed at random and has always the same neighbours, makes very hard to induce the correct behaviour in the society.

\section{Discussion and conclusions}
\label{conclu}

From the previous results, one can conclude that the presence in a society of some agents with infinite persuasion (the monitors) seems to be more effective than to have a group of agents with infinite adamancy (the core of educated agents). But, as we said above, the social cost should be taken into account to have a more complete picture of the problem. On the one hand, a significant limitation is that monitors (police or highway patrols) need to be trained and paid. Also, the number of them necessary to accomplish the percentage $P$ of monitors per agent that our model indicates, should be multiplied at least by $3$ to consider 8 hours work shifts. 

Even within the limitations of a toy model, it is worth comparing our results with a few data from real societies. Our model has several parameters that have not been fitted to data, such as the ratio between the values of the opinions, the size of the influence group, or the distribution of adamancy in the population. Some of them may indeed be very difficult to quantify precisely. Nevertheless we have explored a range of values for these (not shown in this paper) to test for the robustness of the analysis. Even if it has anecdotal value, let us check some statistics about the number of policemen or traffic control agents, which are available on-line. For example, the average number of agents per $100,000$ habitants is $207$ in Canada and the US and 307 in Western Europe~\cite{ONU}. In Porto Alegre, Brazil, the City Council intends to have $700$ traffic agents by $2014$, year of the FIFA World Coup~\cite{ZH}. This figure represents $100$ agents every $100,000$ automobiles. The numbers we found in our simplified model, necessary to reduce the ``wrong behaviour'' to very low levels, is of the order of one monitor every $100$ to $1,000$ agents, i.e. $100$ to $1,000$ every $100,000$ agents, well within the ONU statistics for North America or Western Europe, and also for Porto Alegre projects of traffic control. 

It is also important to notice that monitors in real life do not work twenty four hours a day, nor do they behave as incorruptible agents with infinite persuasion. In Argentina, for example, several major cities may have about 500 policemen per $100,000$ inhabitants (unofficial figures \cite{lavoz}) but they are unable to reduce offences against persons and properties, with and without violence, to levels tolerated by the society.

On the other hand, it is difficult to evaluate the number of educated people, even in simple cases like the respect of traffic rules. The education of a group of the society to form the ``hard nucleus" of the model may well be easier and cheaper, and the effect may be more sustained in time. Certainly a realistic situation should include both approaches: monitors and education. However, it is interesting to remark that the effect of the educated group is much bigger when agents interact at random. This means that social mobility and exchanges with different people are very enriching and favor the diffusion and adoption on the more valued social attitudes. When the educated agents are in fixed locations within a network, their action is less effective and almost unable to induce majors changes with reasonable numbers for the fraction of the educated population. A rigid and limited network of acquaintances can impede or delay the adoption of them.

In any case, it is evident from our results that the combination of both, monitors and educated groups, can change the social behaviour in reasonable lapses of time, and also that the results are improved when the social interactions exhibit dynamic interactions, allowing the agents to observe and compare their behaviour with a large fraction of the society.    

\begin{acknowledgement}
This work received support from the Brazilian Conselho Nacional de Desenvolvimento Científico e Tecnológico (grant CNPq PROSUL-490440/2007), the Argentinian  Consejo Nacional de Investigaciones Científicas y Técnicas (PIP 112-200801-00076), and Universidad Nacional de Cuyo (06/C304). JRI thanks the Argentine Ministry of Science, Technology and Productive Innovation for a Milstein Fellowship that enabled his stay in Bariloche during the last months of 2011, and also CNPq.
\end{acknowledgement}

\end{document}